# Fractal Entropies from the Second Inverse Maximum Entropy Principle

Marcelo R. Ubriaco*

Laboratory of Theoretical Physics
Department of Physics
University of Puerto Rico
Río Piedras Campus
San Juan
PR 00931, USA

**Abstract**

With use of the Second Inverse Maximum Entropy Principle we find entropy functions for systems with fractal distribution functions with order parameter $q$. We compare these entropy functions with those given by the Bose-Einstein and Fermi-Dirac cases.

## 1 Introduction

The concept of entropy in the context of fractals has called a lot of attention, some examples are the study of the entropy : in fractal systems [1], fractal phase space [2], fractal dimensions [3] and fractal geometry [4, 5].
Also it is well known, there has been considerable amount of work to study the consequences of distributions for the average number of particles with energy $\epsilon$ given by [6]

$$< n(\epsilon) >= \frac{1}{[1+\beta(q-1)(\epsilon-\mu)]^{1/(q-1)}} + a, \qquad (1)$$

where $a = 0$ for the classical case, and the values $a = -1$ and $a = 1$ correspond to the non-extensive versions of Bose-Einstein and Fermi-Dirac cases, respectively. For $q = 1$, Equation (1) becomes the standard textbook result for classical and quantum ideal gases. Some of the motivations involving the interest on these distributions are their relation to the theory of fractals,[7], and the fact that their arise as the result of imposing, [8], as an approximation a factorized partition function to the formalism of non-extensive statistical mechanics originally proposed in [9]. This type of approximation has been shown [10] to be good outside a temperature interval that shifts to higher values of $T$ when the number of energy levels increases. One of the interesting results involving these systems is that they are more stable that the standard $q = 1$ ones [11]. In this manuscript we wish to find the entropic functions that would lead to the distributions in Equation (1) subjected to the appropriate constraints. For this purpose we will use the Second Inverse Maximum Entropy Principle (SIMEP), [12] [13], which consists in starting with the required distribution to find the measure of entropy that would lead to the given distribution as a maximum entropy one. This manuscript is organized as follows. In section 2 we briefly describe the formalism of systems with fractal distribution functions as reported in [14]. In Section 3 we describe the method to find the entropy functions from the given distribution subjected to the constraints given by the formalism in Section 2 and compare our results with the standard, $q = 1$, results for Bose-Einstein and Fermi-Dirac statistics. In Section 4 we discuss some mathematical properties of the entropy. In Section 5, we summarize our results.

---

*Electronic address:marcelo.ubriaco@upr.edu

## 2  Fractal Models

The formalism that leads to the distributions in Equation (1) was developed in [14]. In the classical case one defines

$$\rho = \frac{1}{Z_{MB}} \prod_{j=0} \frac{1}{n_j!} \rho_j^{-n_j}, \tag{2}$$

where the function $\rho_j$ is given by

$$\rho_j = [1 + (q-1)\beta(\epsilon_j - \mu)]^{1/(q-1)}, \tag{3}$$

leading after summing over the occupation numbers to the partition function

$$Z_{MB} = \prod_{j=0} e^{(1/\rho_j)}. \tag{4}$$

For the case of Bose-Einstein (BE) and Fermi-Dirac (FD) we found that the probability density is given by the function

$$\rho = \frac{1}{Z} \prod_{j=0} \rho_j^{-n_j}, \tag{5}$$

and the corresponding partition functions can be written as

$$Z = \prod_{j=0} (1 + a\rho_j^{-1})^a, \tag{6}$$

where $a = -1$ for BE and $a = 1$ for FD. In particular, consistency with the classical result $<\epsilon> = \frac{3}{2} <N> kT$ indicates that the correct way of defining the average energy is through the equation

$$<\epsilon> = \frac{4\pi V}{h^3} \int_0^\infty \frac{p^2}{2m} <n(p)>^q p^2 dp. \tag{7}$$

# 3  Second Inverse Maximum Entropy Principle

The SIMEP is used when linear constraints are considered. However, we see that Equation (7) leads to a nonlinear constraint for $q \neq 1$, requiring a more general although almost equally simple approach. We write the unknown entropy measure as

$$S = \int s(<n(\epsilon)>)\, d\epsilon, \tag{8}$$

subjected to the constraints

$$\lambda_1 = \int (<n(\epsilon)>)\, d\epsilon, \tag{9}$$

$$\lambda_2 = \int (<n(\epsilon)>)^q \epsilon\, d\epsilon. \tag{10}$$

Maximizing the unknown measure subjected to these constraints leads to the Equation

$$s' = \lambda_1 + \lambda_2 q \epsilon <n>^{q-1}, \tag{11}$$

where $s' = \frac{ds}{d<n>}$. After taking a second derivative we simply obtain

$$s'' = \frac{q\lambda_2}{<\dot{n}>} <n>^{q-1} + \lambda_2 q(q-1)\epsilon <n>^{q-2}, \tag{12}$$

with $<\dot{n}> = \frac{d<n>}{d\epsilon}$. In particular, for $q = 1$ and identifying $\lambda_2 = \beta$, we have the equations

$$s'' = \frac{\lambda_2}{<\dot{n}>}, \tag{13}$$

$$<n> = \frac{1}{e^{\beta(\epsilon-\mu)} + a}, \tag{14}$$

$$<\dot{n}> = -\beta(<n> - a <n>^2), \tag{15}$$

leading to the expected result

$$S = -\int \left( <n> \ln <n> + \frac{1}{a}(1 - a<n>) \ln(1 - a<n>) \right) + constraints, \tag{16}$$

for Bose-Einstein and Fermi-Dirac statistics. Similarly, if we take a distribution satisfying

$$<\dot{n}> = -\beta <n>^q, \tag{17}$$

we find that

$$s = -\frac{<n>^q}{q-1} + c_1 <n>, \tag{18}$$

such that choosing the integration constant $c_1 = \frac{1}{q-1}$ leads to a Tsallis type entropy with the correct $q \to 1$ limit: $s = - <n> \ln <n>$. For $q \neq 1$ we have from Equation (1)

$$<\dot{n}> = -\beta <n>^q (1 - a<n>)^{2-q}, \tag{19}$$

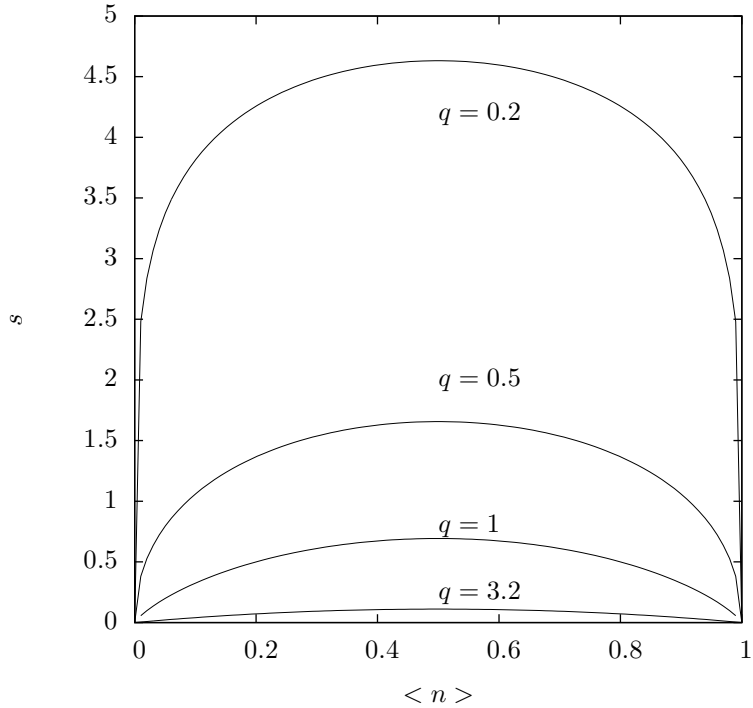

Figure 1: The function $s$, in Equation 21 for $a = 1$, for the case of FD statistics and $q = 0.2, 0.5, 3.2$ in comparison with the case of $q = 1$.

such that Equation (12) becomes

$$s'' = -\frac{q\lambda_2}{\beta}\left(\frac{a}{[1 - a <n>]^{2-q}} + \frac{1}{<n>^{2-q}}\right). \quad (20)$$

After an elementary integration and comparison with Equation (11) we identify the integration constant with $-\lambda_1$ and $\lambda_2 = \beta/q$. Another elementary integration leads to the result

$$s = -\frac{1}{q(q-1)}\left(\frac{1}{a}[1 - a <n>]^q + <n>^q\right) - \lambda_1 <n> + c_2, \quad (21)$$

such that agreement with Equation (16) is achieved in the $q \to 1$ limit by taking the constant $c_2 = \frac{1}{aq(q-1)}$. The entropy then becomes

$$S = -\int \frac{1}{aq(q-1)}\left([1 - a <n>]^q + a <n>^q - 1\right)d\epsilon - \lambda_1 \int <n> d\epsilon. \quad (22)$$

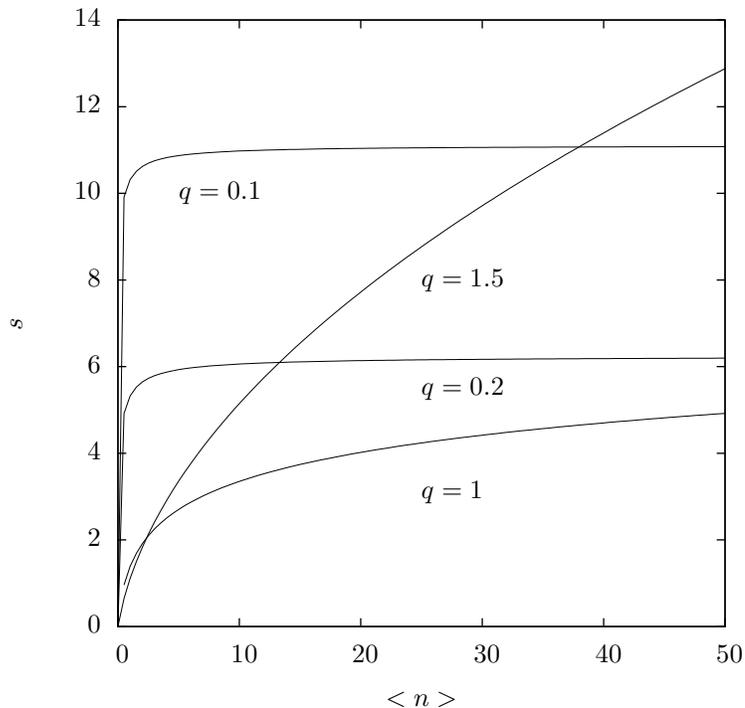

Figure 2: The function $s$, in Equation 21 for $a = -1$, for the case of BE statistics for $q = 0.1, 0.2, 1.5$ in comparison with the case of $q = 1$.

## 4 Mathematical properties

In this Section we consider three basic properties of the function $s$: concavity, positivity and the fact it becomes maximum when all states are equally occupied.

**a** Concavity

From Equation (21)] we write

$$\frac{d^2 s}{d<n>^2} = -\frac{1}{[1-<n>]^{2-q}} - \frac{1}{<n>^{2-q}}, \ (FD) \qquad (23)$$

$$\frac{d^2 s}{d<n>^2} = \frac{1}{[1+<n>]^{2-q}} - \frac{1}{<n>^{2-q}}, \ (BE) \qquad (24)$$

where we find that $s'' < 0 \ \forall q$ for the FD case, and $s'' < 0$ for $0 < q < 2$ for the BE case. In particular, for the BE case at $q = 2$, the function $s \propto <n>$.

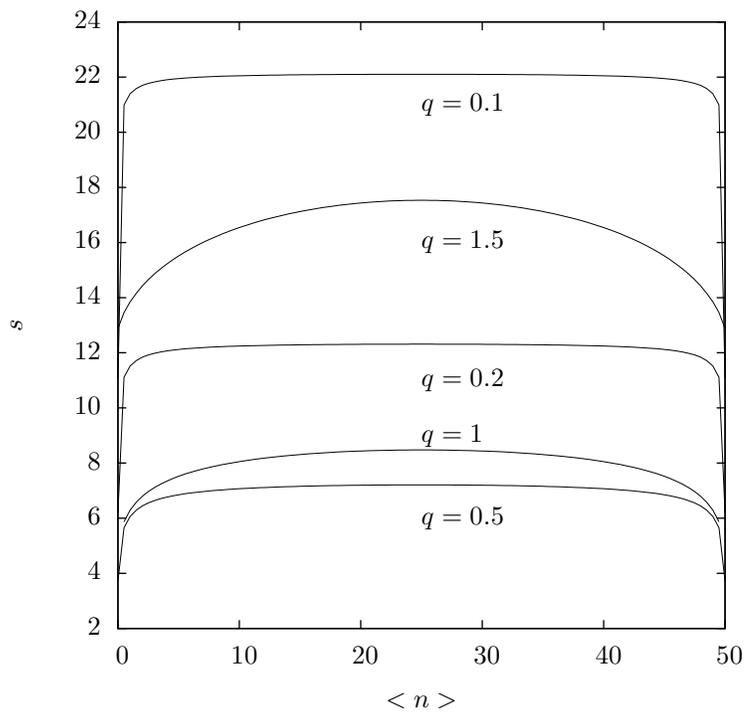

Figure 3: The entropy function $s$ for a two-state system, in Equation 21 for $a = -1$, for the case of BE statistics and $q = 0.1, 0.2, 0.5, 1.5$ and the standard case of $q = 1$.

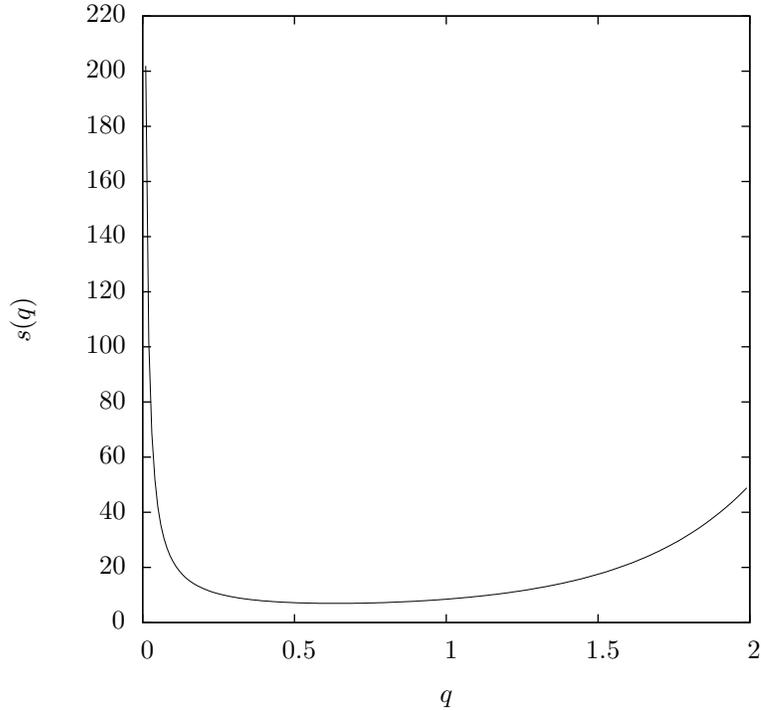

Figure 4: The entropy function $s$ in Equation 21 as a function of the order parameter $q$ for a two-state system for the case of BE statistics and equal occupation number $<n>= 25$.

**b** Positivity

For FD statistics the function $s$ vanishes at $<n>= 0$ and $<n>= 1$. In addition, since the term $(1-<n>)^q + <n>^q -1$ is negative (positive) for $q > 1$ ($q < 1$), the function $s > 0$ $\forall q$.
For BE statistics, the function $s$ can be written as

$$s = \frac{1}{q(q-1)} <n>^q \left( \sum_{l=0} \binom{q}{l} \frac{1}{<n>^l} - 1 - \frac{1}{<n>^q} \right), \quad (25)$$

leading to

$$s = \frac{1}{q(q-1)} \left( q<n>^{q-1} + \frac{q(q-1)}{2!}<n>^{q-2} + \frac{q(q-1)(q-2)}{3!}<n>^{q-3} +... - 1 \right), \quad (26)$$

and the leading term for $1 < q < 2$ as $<n> \to \infty$ is $<n>^{q-1}$, implying therefore that as $<n> \to \infty$ the function $s \to \infty$ for $1 < q < 2$, and $s \to -\frac{1}{q(q-1)}$ for $0 < q < 1$. A very similar inspection tell us that the derivative $\frac{ds}{d<n>} > 0 \ \forall n$, and as $<n> \to \infty$ the derivative $\frac{ds}{d<n>} \to \infty$ for $1 < q < 2$ and $\frac{ds}{d<n>} \to 0$ for $0 < q < 1$.

**c** In order to check that the entropy $S$ is a maximum when all states are equally populated is sufficient to check the case of $M$ particles in two possible states. We simply write

$$S = [1-a<n>]^q + a<n>^q - 1 + [1-a(M-<n>)]^q + a(M-<n>)^q - 1, \tag{27}$$

such that after replacement of $<n> = M/2$ we find that the derivative $S' = 0$, and since $S$ is positive the vanishing derivative corresponds to a maximum.

# 5 Conclusions

In this manuscript with use of the Second Inverse Maximum Entropy Principle we found entropy for systems with fractal distribution functions with order parameter $q$. For the Fermi-Dirac case the fractal entropy $s$ is concave ($s'' < 0$) and positive $\forall q$. For the Bose-Einstein case, $s'' < 0$ for $0 < q < 2$, $s \propto <n>^{q-1}$ for $1 < q < 2$ and $s \to -\frac{1}{q(q-1)}$ as $<n> \to \infty$ for $0 < q < 1$. As expected, the entropy is maximum when all states are equally populated